\documentclass{article}
\usepackage{spconf,graphicx}
\usepackage{amsmath,amssymb, fixmath}
\usepackage{booktabs}
\usepackage{cite}
\usepackage{url,hyperref}
\usepackage{color}

\usepackage{subfig}
\usepackage{float}
\usepackage[dvips]{epsfig}
\usepackage{multirow}

\DeclareMathOperator{\E}{\mathbb{E}}
\DeclareMathOperator{\bx}{\mathbold{x}}
\DeclareMathOperator{\bz}{\mathbold{z}}
\DeclareMathOperator{\bxhat}{\mathbold{\hat{x}}}

\newcommand{\ryuichiedit}[1]{\textcolor{black}{#1}}
\newcommand{\eunwooedit}[1]{\textcolor{black}{#1}}
\newcommand{\redit}[1]{\textcolor{black}{#1}}
\newcommand{\eedit}[1]{\textcolor{black}{#1}}
\newcommand{\Rreviewedit}[1]{\textcolor{black}{#1}}
\newcommand{\Ereviewedit}[1]{\textcolor{black}{#1}}

\newcommand{\waveganmos}{$4.06$ }
\newcommand{\waveganmostts}{$4.16$ }

\title{Parallel WaveGAN: A fast waveform generation model based on generative adversarial networks with multi-resolution spectrogram}

\name{
Ryuichi Yamamoto$^1$, Eunwoo Song$^2$ and Jae-Min Kim$^2$
}
\address{
  $^1$LINE Corp., Tokyo, Japan.\\
  $^2$NAVER Corp., Seongnam, Korea}

\begin{document}
\fontsize{9.2}{11.3}\selectfont
\maketitle
\begin{abstract}
We propose Parallel WaveGAN, a distillation-free, fast, and small-footprint waveform generation method using a generative adversarial network.
In the proposed method, a non-autoregressive WaveNet is trained by jointly optimizing multi-resolution spectrogram and adversarial loss functions, which can effectively capture the time-frequency distribution of the realistic speech waveform.
\Rreviewedit{As our method does not require density distillation used in the conventional teacher-student framework, the entire model can be easily trained. Furthermore, our model is able to generate high-fidelity speech even with its compact architecture.}
In particular, the proposed Parallel WaveGAN has only 1.44 M parameters and can generate 24 kHz speech waveform 28.68 times faster than real-time on a single GPU environment.
Perceptual listening test results verify that our proposed method achieves \waveganmostts mean opinion score within a Transformer-based text-to-speech framework, which is comparative to the best distillation-based Parallel WaveNet system.
\end{abstract}
\section{Introduction}
\label{sec:introduction}

Deep generative models in text-to-speech (TTS) frameworks have significantly improved the quality of synthetic speech signals \cite{ze2013statistical, song2017effective, Shen2018NaturalTS}. 
\ryuichiedit{Remarkably, autoregressive generative models such as WaveNet have shown much superior performance over traditional parametric vocoders \cite{oord2016wavenet, Wang2018WNComparison, tamamori2017speaker,hayashi2017multi, song2019excitnet}. However, they suffer from slow inference speed due to their autoregressive nature and thus are limited in their applications to real-time scenarios.}

One approach to address the limitation is to utilize fast waveform generation methods based on \eunwooedit{a} teacher-student framework \cite{Oord2018ParallelWF, Ping2018ClariNetPW, Yamamoto2019}.
In this framework, a bridge defined as probability density distillation transfers the knowledge of an autoregressive teacher WaveNet to an inverse autoregressive flow (IAF)-based student model \cite{Kingma2016ImprovingVI}. 
Although the IAF student can achieve real-time generation of speech with reasonable perceptual quality, there remain problems in \redit{the} training process: \eunwooedit{it requires not only} a well trained teacher model, but also \ryuichiedit{a trial and error methodology} to optimize the complicated density distillation process.

To overcome the aforementioned problems, we propose a \textit{Parallel WaveGAN}\footnote{
\ryuichiedit{Note that our work is not closely related to an unsupervised waveform synthesis model, WaveGAN~\cite{donahue2018adversarial}.}}, a simple and effective parallel waveform generation method based on \eedit{a} generative adversarial network (GAN) \cite{goodfellow2014generative}. 
Unlike the conventional distillation-based methods, the Parallel WaveGAN does not require the two-stage, sequential teacher-student training process.
In the proposed method, only a non-autoregressive WaveNet model is trained by optimizing the combination of multi-resolution short-time Fourier transform (STFT) and adversarial loss functions that enable the model to effectively capture the time-frequency distribution of the realistic speech waveform.
As a result, the entire training process becomes much easier than the conventional methods, \redit{as well as} the model can produce natural sounding speech waveforms with \ryuichiedit{a} small number of model parameters.
Our contributions are summarized as follows:
\begin{itemize}
   \item We propose a joint training method of the multi-resolution STFT loss and the waveform-domain adversarial loss. This approach effectively works for the conventional distillation-based Parallel WaveNet (e.g., ClariNet), \ryuichiedit{as well as for} the proposed distillation-free \ryuichiedit{Parallel} WaveGAN. 
   \item As the proposed \eunwooedit{Parallel} WaveGAN can be simply trained without any teacher-student framework, our approach significantly reduces both the training and inference time. In particular, the training process becomes 4.82 times faster (from 13.5 days to 2.8 days with two NVIDIA Telsa V100 GPUs) and the inference process becomes 1.96 times faster (from 14.62 to 28.68 real-time\footnote{
   \eunwooedit{The inference speed defined as $k$ means that the system can generate waveforms $k$ times faster than real-time.}} to generate 24 kHz speech waveforms with a single NVIDIA Telsa V100 GPU) compared with the conventional ClariNet model.
   \item We combined the proposed \eunwooedit{Parallel} WaveGAN with a TTS acoustic model based on a Transformer \cite{vaswani2017attention,li2019close,ren2019fastspeech}. 
   \ryuichiedit{The perceptual listening tests verify that the proposed Parallel WaveGAN achieves \waveganmostts MOS, which is competitive to the best distillation-based ClariNet model.}

\end{itemize}

\section{Related work}
\label{sec:related work}

The idea of using GAN in the Parallel WaveNet framework is not new.
In our previous work, the IAF student model was incorporated as a generator and jointly optimized by minimizing the adversarial loss along with the Kullback-Leibler divergence (KLD) and auxiliary losses \cite{Yamamoto2019}.
As the GAN learns the distribution of realistic speech signals,
\ryuichiedit{the method significantly improves the perceptual quality of synthetic signal. However, the \eedit{complicated} training stage based on density distillation limits its utilization.}

Our aim is to minimize the effort to train the two-stage pipeline of \redit{the} conventional teacher-student framework.
In other words, we propose a \ryuichiedit{novel} method to train the Parallel WaveNet without any distillation process.
Juvela et al \cite{Juvela2019} has proposed a similar approach (e.g., GAN-excited linear prediction; GELP) that generates glottal \redit{excitations} by using the adversarial training method.
However, \ryuichiedit{since GELP requires linear prediction (LP) parameters to convert glottal excitations to speech waveform, quality degradation may occur when the LP parameters contain inevitable errors caused by the TTS acoustic model.} 
To avoid this problem, our method is designed to directly estimate the speech waveform. 
As it is very difficult to capture the dynamic nature of speech signal including both the vocal cord movement and the vocal tract resonance (represented by glottal \redit{excitations} and LP parameters in GELP, respectively), we propose a joint optimization method between the adversarial loss and multi-resolution STFT loss in order to capture the time-frequency distributions of the realistic speech signal.
As a result, the entire model can be easily trained even with a small number of parameters while effectively reducing the inference time and improving perceptual quality of synthesized speech.

\section{Method}
\label{sec:method}

\subsection{Parallel waveform generation based on GAN}
\label{sec:parallelgen}

\ryuichiedit{GANs are generative models that are composed \eedit{of} two separate neural networks: a generator ($G$) and a discriminator ($D$)~\cite{goodfellow2014generative}}.
In our method, a WaveNet-based model \ryuichiedit{conditioned on an auxiliary feature (e.g., mel-spectrogram)} is used as the generator, which transforms the input noise to the output waveform in parallel. 
\ryuichiedit{The generator differs from the original WaveNet in that}: (1) we use non-causal convolutions instead of causal convolutions; (2) the input is random noise drawn from a Gaussian distribution; (3) \ryuichiedit{the} model is non-autoregressive \ryuichiedit{at} both training and inference \eunwooedit{steps}.

The generator learns a distribution of realistic waveforms by trying to deceive the discriminator to recognize the generator samples as \textit{real}. 
The process is performed by minimizing the adversarial loss\footnote{
Our method adopts least-squares GANs thanks to its training stability \cite{mao2017least, tian2018generative,Bollepalli2017,pascual2017segan}.} ($L_{\mathrm{adv}}$) as follows:
\begin{equation}
    L_{\mathrm{adv}}(G, D) = \E_{\bz \sim N(0,I)}\left[(1 - D(G(\bz)))^2\right],
    \label{eq:adv}
\end{equation}
where $\bz$ denotes the input white noise. 
\ryuichiedit{Note that the auxiliary feature for $G$ is omitted for brevity.}

\eedit{On the other hand, the} discriminator is trained to correctly classify the generated sample as \textit{fake} while classifying the ground truth as \textit{real} using the following optimization criterion:
\begin{equation}
    L_{\mathrm{D}}(G,D) = \E_{\bx \sim p_{\mathrm{data}}}[(1 - D(\bx))^2] + \E_{\bz \sim N(0, I)}[D(G(\bz))^2], \label{eq:dloss}
\end{equation}
where $\bx$ and $p_{\mathrm{data}}$ denote the target waveform and \eedit{its distribution}, respectively.

\subsection{Multi-resolution STFT auxiliary loss}

\begin{figure}[t]
  \centering
  \centerline{\epsfig{figure=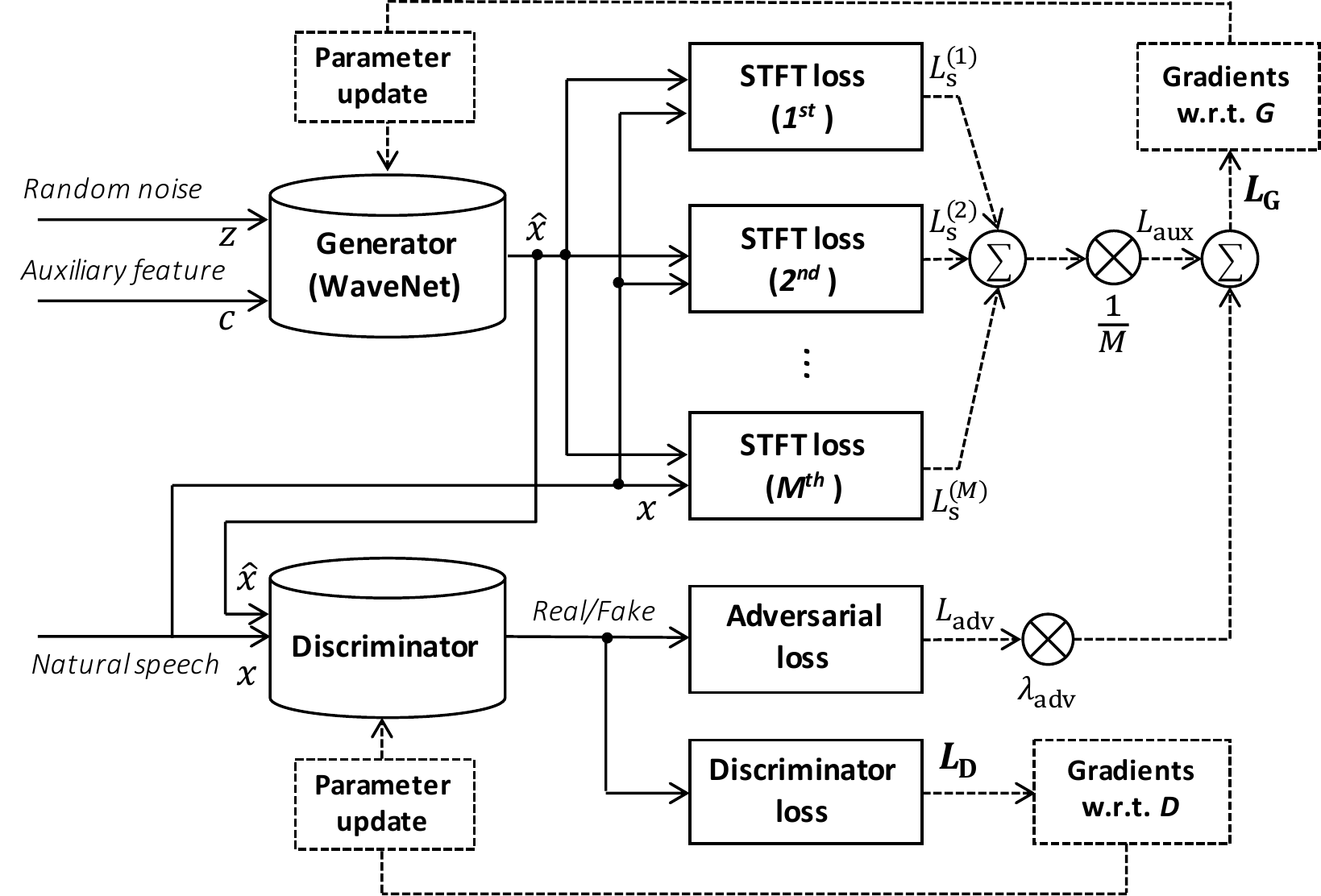,width=1.0\columnwidth}}
  \caption{An illustration of our proposed adversarial training framework with the multi-resolution STFT loss.}
  \label{fig:training}
\end{figure}

\ryuichiedit{To improve the stability and efficiency of the adversarial training} \eedit{process}, we propose a multi-resolution STFT auxiliary loss.
Fig.~\ref{fig:training} shows our framework combining the multi-resolution STFT loss with the adversarial training \eunwooedit{method as} described in section \ref{sec:parallelgen}.

Similar to \eunwooedit{the} previous work~\cite{Yamamoto2019}, we define a \eedit{single} STFT loss as follows:
\begin{equation}
    L_{\mathrm{s}}(G) = \E_{\bz \sim p(\bz), \bx \sim p_{data}} \left[L_{\mathrm{sc}}(\bx,\bxhat) + L_{\mathrm{mag}}(\bx,\bxhat)\right], \label{eq:stftloss}
\end{equation}
where $\bxhat$ denotes the generated sample (i.e., $G(\bz)$), and $L_{\mathrm{sc}}$ and $L_{\mathrm{mag}}$ denote \textit{spectral convergence} and \textit{log STFT magnitude} loss, respectively, which are defined as follows \cite{arik2019fast}:
\begin{equation}
     L_{\mathrm{sc}}(\bx,\bxhat) = \frac{\parallel |\mathrm{STFT}(\bx)| - |\mathrm{STFT}(\bxhat)|\parallel_{F}}{\parallel|\mathrm{STFT}(\bx)|\parallel_{F}},
\end{equation}
\begin{equation}
     L_{\mathrm{mag}}(\bx,\bxhat) = \frac{1}{N}\parallel \mathrm{log}|\mathrm{STFT}(\bx)| - \mathrm{log}|\mathrm{STFT}(\bxhat)| \parallel_{1},
\end{equation}
where $\parallel \cdot \parallel_{F}$ and $\parallel \cdot \parallel_{1}$ denote the Frobenius and $L_1$ norms, respectively; $|\mathrm{STFT}(\cdot)|$ and $N$ denote the STFT magnitudes and number of elements in the magnitude, respectively. 

Our multi-resolution STFT loss is the sum of the \eedit{STFT} losses with different analysis parameters (i.e., FFT size, window size, and frame shift). Let $M$ be the number of STFT losses, the multi-resolution STFT auxiliary loss ($L_{\mathrm{aux}}$) is represented as follows: 
\begin{equation}
    L_{\mathrm{aux}}(G) = \frac{1}{M}\sum_{m=1}^{M}L^{(m)}_{\mathrm{s}}(G). \label{eq:multispec}
\end{equation}
In \eedit{the} STFT-based time-frequency representation of signals, there is a trade-off between time and frequency resolution; e.g., increasing window size gives higher frequency resolution \eedit{while reducing temporal} resolution\cite{daubechies1990wavelet}.
By \ryuichiedit{combining} multiple STFT losses with different analysis parameters, it greatly helps the generator to learn \eedit{the} time-frequency characteristics of speech \cite{wang2019neural}.
Moreover, it also prevents the generator from \eedit{being overfit} to a fixed STFT representation, \ryuichiedit{which may result in suboptimal performance in the waveform-domain.} 

Our final loss function for the generator is \eedit{defined as} a linear combination of the multi-resolution STFT loss and the adversarial loss as follows:
\begin{equation}
    L_{\mathrm{G}}(G, D) = L_{\mathrm{aux}}(G) + \lambda_{\mathrm{adv}}L_{\mathrm{adv}}(G, D), \label{eq:gloss}
\end{equation}
where $\lambda_{\mathrm{adv}}$ denotes the hyperparameter balancing the two loss terms.
\ryuichiedit{By jointly optimizing the waveform-domain adversarial loss and the multi-resolution STFT loss, the generator can learn the distribution of the realistic speech waveform effectively.}


\section{Experiments}
\label{sec:experiments}

\subsection{Experimental setup}

\subsubsection{Database}

In the experiments, we used a phonetically and prosaically balanced speech corpus recorded by a female professional Japanese speaker. 
The speech signals were sampled at 24 kHz, each sample was quantized by 16 bits. 
In total, 11,449 utterances (23.09 hours) were used for training, 250 utterances (0.35 hours) were used for validation, and another 250 utterances (0.34 hours) were used for evaluation.
The 80-band log-mel spectrograms with band-limited frequency range\footnote{\ryuichiedit{We empirically found that using the band-limited features alleviates the over-smoothing problem caused by acoustic models in TTS.}} (70 to 8000 Hz) were extracted and used \ryuichiedit{as the input auxiliary features for waveform generation models (i.e., local-conditioning \cite{oord2016wavenet}).}
The frame and shift lengths were set to 50 ms and 12.5 ms, respectively. 
The mel-spectrogram features were normalized to have zero mean and unit variance before training.

\subsubsection{Model details}

The proposed Parallel WaveGAN consisted of 30 layers of dilated residual convolution blocks with exponentially increasing three dilation cycles \cite{oord2016wavenet}.
The number of residual \eedit{and} skip channels \redit{were} \eedit{set to} 64 and \ryuichiedit{the convolution} filter size was \eedit{set to} \eunwooedit{three}. 
The discriminator consisted of ten layers of non-causal dilated 1-D convolutions with leaky ReLU activation function ($\alpha = 0.2$). 
The strides were set to \eunwooedit{one} and linearly increasing dilations were applied for the 1-D convolutions starting from \eunwooedit{one to eight} except for the first and last layers. \ryuichiedit{The number of channels and filter size were the same as the generator}. 
\redit{We applied weight normalization to all \redit{convolutional} layers for both the generator and the discriminator~\cite{salimans2016weight}.}

At the training stage, the multi-resolution STFT loss was computed by the sum of three different STFT losses as described in Table~\ref{tab:stft}. The discriminator loss was computed by the average of per-time step scalar predictions with the discriminator.
The hyperparameter $\lambda_{\mathrm{adv}}$ in equation~(\ref{eq:gloss}) was chosen to be 4.0 based on our preliminary experiments.
Models were trained for 400 K steps with RAdam optimizer ($\epsilon=1e^{-6}$) to stabilize training~\cite{liu2019radam}. Note that the discriminator was fixed for the first 100K steps, and two models were jointly trained afterwards.
The minibatch size was \eedit{set to} eight and the length of each audio clip was \eedit{set to} 24 K time samples (\eunwooedit{1.0 second}).
The initial learning rate was set to $0.0001$ and $0.00005$ for the generator and discriminator, respectively. The learning rate was reduced by half for every 200 K steps.

As baseline systems, we used both the autoregressive \ryuichiedit{Gaussian} WaveNet and the parallel one (\redit{i.e.}, ClariNet)~\cite{Ping2018ClariNetPW,Yamamoto2019}.
The WaveNet consisted of 24 layers of dilated residual convolution blocks with four dilation cycles. 
The number of residual \eedit{and} skip channels were \eedit{set to} 128 and the filter size was \eedit{set to} \eunwooedit{three}. 
The model was trained for 1.5 M steps with RAdam optimizer. 
The learning rate was set to $0.001$, and it was reduced by half for every 200 K steps.
The minibatch size was \eedit{set to} eight and the length of each audio clip was \eedit{set to} 12 K time samples (\eunwooedit{0.5 second}). 

To train the baseline ClariNet, the autoregressive WaveNet described above was used as the teacher model.
The ClariNet was based on Gaussian IAFs \cite{Ping2018ClariNetPW}, \ryuichiedit{which} consisted of six flows. 
Each flow was parameterized by ten layers of dilated residual convolution blocks with an exponentially increasing dilation cycle. 
The number of residual \eedit{and} skip channels \redit{were} \eedit{set to} 64 and the filter size was \eedit{set to} \eunwooedit{three}.
The weight coefficients to balance the KLD and STFT auxiliary losses were set to 0.5 and 1.0, respectively.
The model was trained for 400 K steps with the same optimizer settings \eedit{of} the Parallel WaveGAN. 
\ryuichiedit{We also investigated ClariNet with adversarial loss as a hybrid approach of GAN and density distillation \cite{Yamamoto2019}}. The model structure was the same as the baseline ClariNet, but it was trained with the mixture of KLD, STFT and adversarial losses, where the weight coefficients to balance them were set to 0.05, 1.0 and 4.0, respectively. The model was trained for 200K steps with the fixed discriminator, and the generator and discriminator were jointly \ryuichiedit{optimized} for the rest 200 K steps.

Throughout the waveform generation models, the input auxiliary features were upsampled by nearest neighbor upsampling followed by 2-D convolutions \redit{so that the time-resolution of auxiliary features matches the sampling rate of the speech waveform~\cite{odena2016deconvolution,Yamamoto2019}}.
Note that the auxiliary features were not used for discriminators.
All the models were trained using two NVIDIA Tesla V100 GPUs. Experiments were conducted on the NAVER smart machine learning (NSML) platform \cite{kim2018nsml}.

\begin{table}[t]
  \caption{The details of the multi-resolution STFT loss. A Hanning window was applied before the FFT process.}
  \vspace{-2mm}
  \label{tab:stft}
  \centering
  \scalebox{0.90}{
  {\renewcommand\arraystretch{0.90}
  \begin{tabular}{cccc}
    \toprule
    \textbf{STFT loss} & \textbf{FFT size} & \textbf{Window size} & \textbf{Frame shift}\\
    \midrule
    $L_{\mathrm{s}}^{(1)}$ & 1024 & 600 (25 ms) & 120 (5 ms) \\
    $L_{\mathrm{s}}^{(2)}$ & 2048 & 1200 (50 ms) & 240 (10 ms) \\
    $L_{\mathrm{s}}^{(3)}$ & 512 & 240 (10 ms) & 50 ($\approx$ 2 ms) \\
    \bottomrule
  \vspace{-6mm}
  \end{tabular}
  }
  }
\end{table}

\begin{table*}[t]
  \caption{The inference speed and the MOS results with 95\% confidence intervals: Acoustic features extracted from the recorded speech signal were used to compose the input auxiliary features. The evaluation was conducted on a server with a single NVIDIA Tesla V100 GPU. Note that the inference speed \eunwooedit{${k}$} means that the system was able to generate waveforms \eunwooedit{${k}$} times faster than real-time.}
  \vspace{-2mm}
  \label{tab:systems}
  \centering
  \scalebox{0.80}{
  {\renewcommand\arraystretch{0.80}

  \begin{tabular}{ccccccccc}
    \toprule
    \textbf{System} & \multirow{2}*{\textbf{Model}} & \textbf{KLD-based} & \textbf{STFT} & \textbf{Adversarial} & \textbf{Number of} & \textbf{Model} & \textbf{Inference} & \multirow{2}*{\textbf{MOS}} \\
    \textbf{index} & & \textbf{distillation} & \textbf{loss} & \textbf{loss} & \textbf{layers} & \textbf{size} & \textbf{speed} \\
    \midrule
    System 1 & WaveNet & - & - & - & 24 & 3.81 M & 0.32$\times10^{-2}$ & 3.61$\pm$0.12 \\
    System 2 & ClariNet & Yes & $L_{\mathrm{s}}^{(1)}$ & - & 60 & 2.78 M & 14.62 & 3.88$\pm$0.11 \\
    System 3 & ClariNet & Yes & $L_{\mathrm{s}}^{(1)}+L_{\mathrm{s}}^{(2)}+L_{\mathrm{s}}^{(3)}$ & - & 60 & 2.78 M & 14.62 & 4.21$\pm$0.09 \\
    System 4 & ClariNet & Yes & $L_{\mathrm{s}}^{(1)}+L_{\mathrm{s}}^{(2)}+L_{\mathrm{s}}^{(3)}$ & Yes & 60 & 2.78 M & 14.62 & 4.21$\pm$0.09 \\
    System 5 & \ryuichiedit{Parallel WaveGAN} & - & $L_{\mathrm{s}}^{(1)}$ & Yes & 30 & 1.44 M & 28.68 & 1.36$\pm$0.07 \\
    System 6 & \ryuichiedit{Parallel WaveGAN} & - & $L_{\mathrm{s}}^{(1)}+L_{\mathrm{s}}^{(2)}+L_{\mathrm{s}}^{(3)}$ & Yes & 30 & 1.44 M & 28.68 & 4.06$\pm$0.10 \\
    \midrule
    System 7 & Recording & - & - & - & - & - & - & 4.46$\pm$0.08 \\
    \bottomrule
    \vspace{-7mm}
  \end{tabular}
  }
  }
\end{table*}

\subsection{Evaluation}
\label{ssec:ch4-2}

To evaluate the perceptual quality, \ryuichiedit{we performed} mean opinion score (MOS)\footnote{Audio samples are available at the following URL:\\\url{https://r9y9.github.io/demos/projects/icassp2020/}} tests.
Eighteen native Japanese speakers were asked to make quality judgments about the synthesized speech samples using the following five possible responses: 1 = Bad; 2 = Poor; 3 = Fair; 4 = Good; and 5 = Excellent. 
In total, 20 utterances were randomly selected from the \eedit{evaluation} set and were then synthesized using the different models. 

Table~\ref{tab:systems} shows the inference speed and the MOS test results with respect to different generation models.
The findings can be summarized as follows: (1) the systems trained with the STFT loss performed better than ones trained without the STFT loss (i.e., the autoregressive WaveNet).
Note that most listeners were unsatisfied with the high-frequency noise caused by the autoregressive WaveNet system.
This could be explained by the fact that only the band-limited (70 - 8000 Hz) mel-spectrogram was used for local-conditioning \eedit{in} the WaveNet, while the other systems were able to directly learn full-band frequency information via STFT loss. 
(2) The proposed multi-resolution STFT loss-based models showed higher perceptual quality than the conventional single STFT losss-based ones (comparing System 3 and 6 with System 2 and 5, respectively).
This confirms that the multi-resolution STFT loss effectively captured the time-frequency characteristics of the speech signal, enabling to achieve better performance.
(3) The proposed adversarial loss did not work well with the ClariNet. 
However, its advantage could be found when it was combined with a TTS framework, which will be discussed in the next section.
(4) Finally, the proposed Parallel WaveGAN achieved \waveganmos MOS. 
Although its perceptual quality was relatively worse than the ClariNet's, the Parallel WaveGAN was able to generate speech signal $1.96$ times faster than the ClariNet.
Furthermore, the benefit of the proposed method could be found in its simple training procedure.
We \ryuichiedit{measured} the total training time for obtaining the optimal models, as described in Table \ref{tab:training-time}.
Because the Parallel WaveGAN did not require any complicated density distillation, it only took 2.8 training days to be optimized, which was $2.64$ and $4.82$ times faster than the autoregressive WaveNet and the ClariNet, respectively.

\begin{table}[t]
  \caption{Training time comparison: All the experiments were conducted on a server with two NVIDIA Tesla V100 GPUs. Each vocoder model corresponds to System 1, 3, 4, and 6 described in Table~\ref{tab:systems}, respectively. Note that the times for ClariNets include the training time for the teacher WaveNet.}
  \vspace{-2mm}
  \label{tab:training-time}
  \centering
  \scalebox{0.90}{
  {\renewcommand\arraystretch{0.90}

  \begin{tabular}{lc}
    \toprule
    \textbf{Model} & \textbf{Training time (days)}\\
    \midrule
    WaveNet & 7.4 \\
    ClariNet & 12.7 \\
    ClariNet-GAN & 13.5 \\
    \ryuichiedit{Parallel WaveGAN} (ours)& 2.8 \\
    \bottomrule
  \vspace{-7mm}
  \end{tabular}
  }
  }
\end{table}

\subsection{Text-to-speech}
\label{ssec:ch4-3}

To verify the effectiveness of the proposed method as the vocoder of the TTS framework, we combined the Parallel WaveGAN with the Transformer-based parameter estimator~\cite{vaswani2017attention,li2019close,ren2019fastspeech}.


To train the Transformer, we used the phoneme \redit{sequences} as input and mel-spectrograms extracted from the recorded speech as output.
The model consisted of a \eunwooedit{six}-layer encoder and \eunwooedit{a six}-layer decoder, each was based on multi-head attention (with \eunwooedit{eight} heads). The configuration followed the prior work~\cite{ren2019fastspeech}, but the model was modified to accept accent as \ryuichiedit{an} external input for pitch accent language (e.g., Japanese)~\cite{yasuda2019investigation}. The model was trained for 1000 epochs using RAdam optimizer with warmup learning rate scheduling~\cite{vaswani2017attention}. Initial learning rate was set to 1.0 and dynamic batch size (average 64) strategy was used to stabilize training.

In the synthesis step, the input phoneme and accent sequences were converted to the corresponding mel-spectrograms by the Transformer \ryuichiedit{TTS} model. By inputting resulting acoustic parameters, \redit{vocoder models} generated the time-domain speech signal.

To evaluate the quality of the generated speech samples, \ryuichiedit{we performed} MOS tests.
The test setups were the same as those described in section~\ref{ssec:ch4-2}, but \ryuichiedit{we used} the autoregressive WaveNet and the parallel generation models trained with the multi-resolution STFT loss in the test (System 1, 3, 4, and 6 described in Table~\ref{tab:systems}, respectively).
The results of the MOS tests \eedit{are} shown in Table~\ref{tab:mos}, of which findings can be summarized as follows: (1) the ClariNet trained with the adversarial loss performed better than the system trained without the adversarial loss, although their perceptual qualities were almost same in the analysis/synthesis case (System 3 and 4 shown in Table~\ref{tab:systems}).
This implies that the use of adversarial loss was advantageous for improving the model's robustness to the prediction errors caused by the acoustic model.
(2) The merits of the adversarial training were also beneficial to the proposed Parallel WaveGAN system.
Consequently, the Parallel WaveGAN with the Transformer TTS model achieved \waveganmostts MOS, which was comparable to the best distillation-based Parallel WaveNet system (i.e., ClariNet-GAN).

\begin{table}[t]
  \caption{MOS results with 95\% confidence intervals: Acoustic features generated from the Transformer \ryuichiedit{TTS} model were used to compose the input auxiliary features.}
  \vspace{-2mm}
  \label{tab:mos}
  \centering
  \scalebox{0.90}{
  {\renewcommand\arraystretch{0.90}

  \begin{tabular}{lc}
    \toprule
    \textbf{Model} & \textbf{MOS}\\
    \midrule
    Transformer + WaveNet & 3.33$\pm$0.11 \\
    Transformer + ClariNet & 4.00$\pm$0.10 \\
    Transformer + ClariNet-GAN & 4.14$\pm$0.10 \\
    Transformer + \ryuichiedit{Parallel WaveGAN} (ours) & 4.16$\pm$0.09 \\
    \midrule
    Recording & 4.46$\pm$0.08 \\
    \bottomrule
  \vspace{-7mm}
  \end{tabular}
  }
  }
\end{table}

\section{Conclusion}
\label{sec:summary}

We proposed Parallel WaveGAN, a distillation-free, fast, and small-footprint waveform generation method based on GAN. 
By jointly optimizing waveform-domain adversarial loss and multi-resolution STFT loss, our model was able to learn how to generate realistic waveforms without any \ryuichiedit{complicated} probability density distillation.
Experimental results demonstrated that our proposed method \eunwooedit{achieved} \waveganmostts MOS within the \ryuichiedit{Transformer-based} TTS framework competitive to the conventional distillation-based approaches, generating 24 kHz speech waveform 28.68 times faster than real-time with only 1.44 M model parameters.
\Rreviewedit{\Ereviewedit{Future research includes improving} the multi-resolution STFT auxiliary loss to better capture the characteristics of speech (e.g., introducing phase-related loss), \Ereviewedit{and verifying its performance to} a variety of corpora, including expressive ones.}

\section{Acknowledgements}
The work was supported by Clova Voice, NAVER Corp., Seongnam, Korea.
The authors would like to thank Adrian Kim, Jung-Woo Ha, Muhammad Ferjad Naeem, and Xiaodong Gu at NAVER Corp., Seongnam, Korea, for their support.

\bibliographystyle{IEEEbib}
\bibliography{refs}

\begin{thebibliography}{10}
\providecommand{\url}[1]{#1}
\csname url@samestyle\endcsname
\providecommand{\newblock}{\relax}
\providecommand{\bibinfo}[2]{#2}
\providecommand{\BIBentrySTDinterwordspacing}{\spaceskip=0pt\relax}
\providecommand{\BIBentryALTinterwordstretchfactor}{4}
\providecommand{\BIBentryALTinterwordspacing}{\spaceskip=\fontdimen2\font plus
\BIBentryALTinterwordstretchfactor\fontdimen3\font minus
  \fontdimen4\font\relax}
\providecommand{\BIBforeignlanguage}[2]{{%
\expandafter\ifx\csname l@#1\endcsname\relax
\typeout{** WARNING: IEEEtran.bst: No hyphenation pattern has been}%
\typeout{** loaded for the language `#1'. Using the pattern for}%
\typeout{** the default language instead.}%
\else
\language=\csname l@#1\endcsname
\fi
#2}}
\providecommand{\BIBdecl}{\relax}
\BIBdecl

\bibitem{ze2013statistical}
H.~Zen, A.~Senior, and M.~Schuster, ``Statistical parametric speech synthesis
  using deep neural networks,'' in \emph{Proc. ICASSP}, 2013, pp. 7962--7966.

\bibitem{song2017effective}
E.~Song, F.~K. Soong, and H.-G. Kang, ``Effective spectral and excitation
  modeling techniques for {LSTM}-{RNN}-based speech synthesis systems,''
  \emph{IEEE/ACM Trans. Audio, Speech, and Lang. Process.}, vol.~25, no.~11,
  pp. 2152--2161, 2017.

\bibitem{Shen2018NaturalTS}
J.~Shen, R.~Pang, R.~J. Weiss, M.~Schuster, N.~Jaitly, Z.~Yang, Z.~Chen,
  Y.~Zhang, Y.~Wang, R.~Skerrv-Ryan \emph{et~al.}, ``Natural {TTS} synthesis by
  conditioning {W}ave{N}et on mel spectrogram predictions,'' in \emph{Proc.
  ICASSP}, 2018, pp. 4779--4783.

\bibitem{oord2016wavenet}
A.~van~den Oord, S.~Dieleman, H.~Zen, K.~Simonyan, O.~Vinyals, A.~Graves,
  N.~Kalchbrenner, A.~Senior, and K.~Kavukcuoglu, ``{WaveNet}: {A} generative
  model for raw audio,'' \emph{arXiv preprint arXiv:1609.03499}, 2016.

\bibitem{Wang2018WNComparison}
X.~Wang, J.~Lorenzo-Trueba, S.~Takaki, L.~Juvela, and J.~Yamagishi, ``A
  comparison of recent waveform generation and acoustic modeling methods for
  neural-network-based speech synthesis,'' in \emph{Proc. ICASSP}, 2018, pp.
  4804--4808.

\bibitem{tamamori2017speaker}
A.~Tamamori, T.~Hayashi, K.~Kobayashi, K.~Takeda, and T.~Toda,
  ``Speaker-dependent {W}ave{N}et vocoder,'' in \emph{Proc. INTERSPEECH}, 2017,
  pp. 1118--1122.

\bibitem{hayashi2017multi}
T.~Hayashi, A.~Tamamori, K.~Kobayashi, K.~Takeda, and T.~Toda, ``An
  investigation of multi-speaker training for {W}ave{N}et vocoder,'' in
  \emph{Proc. ASRU}, 2017, pp. 712--718.

\bibitem{song2019excitnet}
E.~Song, K.~Byun, and H.-G. Kang, ``Excitnet vocoder: {A} neural excitation
  model for parametric speech synthesis systems,'' in \emph{Proc. EUSIPCO},
  2019, pp. 1179--1183.

\bibitem{Oord2018ParallelWF}
A.~van~den Oord, Y.~Li, I.~Babuschkin, K.~Simonyan, O.~Vinyals, K.~Kavukcuoglu,
  G.~van~den Driessche, E.~Lockhart, L.~C. Cobo, F.~Stimberg \emph{et~al.},
  ``Parallel {W}ave{N}et: Fast high-fidelity speech synthesis,'' in \emph{Proc.
  ICML}, 2018, pp. 3915--3923.

\bibitem{Ping2018ClariNetPW}
W.~Ping, K.~Peng, and J.~Chen, ``{ClariNet}: Parallel wave generation in
  end-to-end text-to-speech,'' in \emph{Proc. ICLR}, 2019.

\bibitem{Yamamoto2019}
R.~Yamamoto, E.~Song, and J.-M. Kim, ``Probability density distillation with
  generative adversarial networks for high-quality parallel waveform
  generation,'' in \emph{Proc. INTERSPEECH}, 2019, pp. 699--703.

\bibitem{Kingma2016ImprovingVI}
D.~P. Kingma, T.~Salimans, R.~Jozefowicz, X.~Chen, I.~Sutskever, and
  M.~Welling, ``Improved variational inference with inverse autoregressive
  flow,'' in \emph{Proc. NIPS}, 2016, pp. 4743--4751.

\bibitem{donahue2018adversarial}
C.~Donahue, J.~McAuley, and M.~Puckette, ``Adversarial audio synthesis,'' in
  \emph{Proc. ICLR}, 2019.

\bibitem{goodfellow2014generative}
I.~Goodfellow, J.~Pouget-Abadie, M.~Mirza, B.~Xu, D.~Warde-Farley, S.~Ozair,
  A.~Courville, and Y.~Bengio, ``Generative adversarial nets,'' in \emph{Proc.
  NIPS}, 2014, pp. 2672--2680.

\bibitem{vaswani2017attention}
A.~Vaswani, N.~Shazeer, N.~Parmar, J.~Uszkoreit, L.~Jones, A.~N. Gomez,
  {\L}.~Kaiser, and I.~Polosukhin, ``Attention is all you need,'' in
  \emph{Proc. NIPS}, 2017, pp. 5998--6008.

\bibitem{li2019close}
N.~Li, S.~Liu, Y.~Liu, S.~Zhao, M.~Liu, and M.~T. Zhou, ``Neural speech
  synthesis with {T}ransformer network,'' in \emph{Proc. AAAI}, 2019, pp.
  6706--6713.

\bibitem{ren2019fastspeech}
Y.~Ren, Y.~Ruan, X.~Tan, T.~Qin, S.~Zhao, Z.~Zhao, and T.~Liu, ``{FastSpeech}:
  Fast, robust and controllable text to speech,'' \emph{arXiv preprint
  arXiv:1905.09263}, 2019.

\bibitem{Juvela2019}
L.~Juvela, B.~Bollepalli, J.~Yamagishi, and P.~Alku, ``{GELP: GAN}-excited
  linear prediction for speech synthesis from mel-spectrogram,'' in \emph{Proc.
  INTERSPEECH}, 2019, pp. 694--698.

\bibitem{mao2017least}
X.~Mao, Q.~Li, H.~Xie, R.~Y. Lau, Z.~Wang, and S.~Paul~Smolley, ``Least squares
  generative adversarial networks,'' in \emph{Proc. ICCV}, 2017, pp.
  2794--2802.

\bibitem{tian2018generative}
Q.~Tian, X.~Wan, and S.~Liu, ``Generative adversarial network based speaker
  adaptation for high fidelity {WaveNet} vocoder,'' in \emph{Proc. SSW}, 2019,
  pp. 19--23.

\bibitem{Bollepalli2017}
B.~Bollepalli, L.~Juvela, and P.~Alku, ``Generative adversarial network-based
  glottal waveform model for statistical parametric speech synthesis,'' in
  \emph{Proc. INTERSPEECH}, 2017, pp. 3394--3398.

\bibitem{pascual2017segan}
S.~Pascual, A.~Bonafonte, and J.~Serrà, ``{SEGAN}: Speech enhancement
  generative adversarial network,'' in \emph{Proc. INTERSPEECH}, 2017, pp.
  3642--3646.

\bibitem{arik2019fast}
S.~{\"O}. Ar{\i}k, H.~Jun, and G.~Diamos, ``Fast spectrogram inversion using
  multi-head convolutional neural networks,'' \emph{IEEE Signal Procees.
  Letters}, vol.~26, no.~1, pp. 94--98, 2019.

\bibitem{daubechies1990wavelet}
I.~Daubechies, ``The wavelet transform, time-frequency localization and signal
  analysis,'' \emph{IEEE trans. on information theory}, vol.~36, no.~5, pp.
  961--1005, 1990.

\bibitem{wang2019neural}
X.~Wang, S.~Takaki, and J.~Yamagishi, ``Neural source-filter-based waveform
  model for statistical parametric speech synthesis,'' in \emph{Proc. ICASSP},
  2019, pp. 5916--5920.

\bibitem{salimans2016weight}
T.~Salimans and D.~P. Kingma, ``Weight normalization: A simple
  reparameterization to accelerate training of deep neural networks,'' in
  \emph{Proc. NIPS}, 2016, pp. 901--909.

\bibitem{liu2019radam}
L.~Liu, H.~Jiang, P.~He, W.~Chen, X.~Liu, J.~Gao, and J.~Han, ``On the variance
  of the adaptive learning rate and beyond,'' \emph{arXiv preprint
  arXiv:1908.03265}, 2019.

\bibitem{odena2016deconvolution}
\BIBentryALTinterwordspacing
A.~Odena, V.~Dumoulin, and C.~Olah, ``Deconvolution and checkerboard
  artifacts,'' \emph{Distill}, 2016. [Online]. Available:
  \url{http://distill.pub/2016/deconv-checkerboard}
\BIBentrySTDinterwordspacing

\bibitem{kim2018nsml}
H.~Kim, M.~Kim, D.~Seo, J.~Kim, H.~Park, S.~Park, H.~Jo, K.~Kim, Y.~Yang,
  Y.~Kim \emph{et~al.}, ``{NSML}: Meet the mlaas platform with a real-world
  case study,'' \emph{arXiv preprint arXiv:1810.09957}, 2018.

\bibitem{yasuda2019investigation}
Y.~Yasuda, X.~Wang, S.~Takaki, and J.~Yamagishi, ``Investigation of enhanced
  {T}acotron text-to-speech synthesis systems with self-attention for pitch
  accent language,'' in \emph{Proc. ICASSP}, 2019, pp. 6905--6909.

\end{thebibliography}

\end{document}